\documentclass[12pt]{article}
\usepackage{graphicx}
\usepackage{graphics}

\newcommand\pubnumber{SNSN-323-63}
\newcommand\pubdate{\today}

\def\padova{Universit\`a di Padova and \\
INFN Padova, ITALY}

\def\Title#1{\begin{center} {\Large #1 } \end{center}}
\def\Author#1{\begin{center}{ \sc #1} \end{center}}
\def\Address#1{\begin{center}{ \it #1} \end{center}}

\newcommand\pubblock{\rightline{\begin{tabular}{l} \pubnumber\\
         \pubdate  \end{tabular}}}
\newenvironment{Abstract}{\begin{quotation}  }{\end{quotation}}
\newenvironment{Presented}{\begin{quotation} \begin{center} 
             PRESENTED AT\end{center}\bigskip 
      \begin{center}\begin{large}}{\end{large}\end{center} \end{quotation}}

\begin{document}
\begin{titlepage}
\pubblock

\vfill
\Title{A neutrino interaction with two vertices topology detected by OPERA}
\vfill
\Author{ Eduardo Medinaceli \\
on behalf of the OPERA collaboration}
\Address{\padova}
\vfill
\begin{Abstract}
The OPERA experiment has reported the detection of five $\nu_{\tau}$ candidates
in the CNGS $\nu_{\mu}$ beam, allowing to reject the background-only hypothesis at the 5.1$\sigma$
level. Besides these events, on May 23$^{\mbox{rd}}$ 2011, OPERA detected a ``neutral current like'' 
interaction with two secondary vertices.
Such topologies mainly arise from Charged Current interactions of a $\nu_{\tau}$ with associated
charm quark production or from Neutral Current interactions of a $\nu_{\mu}$ with
production of a charm anti-charm pair. These topologies have generally low probabilities.
A dedicated multivariate analysis is in progress to allow discriminating 
between these two hypotheses. Here the event topology is described in detail and preliminary 
results of the classifiers for all possible contributions are given.
\end{Abstract}
\vfill
\begin{Presented}
NuPhys2015, Prospects in Neutrino Physics\\
Barbican Centre, London, UK,  December 16--18, 2015
\end{Presented}
\vfill
\end{titlepage}
\def\thefootnote{\fnsymbol{footnote}}
\setcounter{footnote}{0}

\section{The OPERA experiment}
\vspace{-0.2cm}
The OPERA (Oscillation Project with Emulsion tRacking Apparatus) experiment has searched for
neutrino oscillations over a baseline of 730 km. The experiment was exposed to the CNGS 
high energy neutrino beam (mean $E_{\nu_{\mu}}\sim 17$ GeV) optimized for the appearance of 
$\nu_{\tau}$ from a pure beam of $\nu_{\mu}$ \cite{opera_proposal}. 
OPERA was a modular detector composed of two Super Modules (SM) each being composed of a target region 
followed by a muon spectrometer. 
The basic unit of the target was the so-called ``brick'', an array of 57 nuclear emulsion layers 
interleaved with 56 $1$ mm thick lead layers; its transverse size was $12.5\times10.2$ cm$^2$ and the 
thickness was about 10 radiation lengths \cite{detector}. Nuclear emulsion have a sub-micrometric 
spatial resolution which grants the precise reconstruction of the neutrino interaction inside the 
target, as well as the associated short-lived particles like the $\tau$ lepton. An OPERA film has 2 
emulsion layers (each 44 $\mu$m thick) on both sides of a transparent triacetylcellulose base (205 $\mu$m 
thick). The overall target mass was 1.25 kton, composed of about 150000 bricks. Each brick is a stand-alone 
device which allows momentum measurement through Multiple Coulomb Scattering (MCS) in the lead plates 
\cite{coulomb}, and electromagnetic shower energy reconstruction. The target region was instrumented with 
plastic scintillator strips \cite{TT}, used to localize the neutrino interactions point; providing the 
probability map of the bricks where the interaction might have taken place. The muon spectrometers were 
equipped with Resistive Plate Chambers and drift tube detectors. The design leads to a muon momentum 
resolution better than $30\%$ for momenta up to 25 GeV/c. Thanks to the high signal to background ratio the 
$\nu_{\tau}$ search was done on an event by event basis \cite{decaysearch}. This possibility also relies 
on the excellent spatial resolution of nuclear emulsions and on the good knowledge of the expected decay 
topologies; having the capability to identify anomalous impact parameters (IP) 
of particle tracks with respect to the reconstructed primary interaction vertex. This is the trigger for the $\tau$ lepton 
decay signal. The charm decay dataset is used as a proxy to validate the reconstruction method, because of 
the similarity decay topologies. Further background suppression is achieved by applying kinematical selections.

OPERA has reported the detection of five $\nu_{\tau}$ candidates, allowing to reject 
the background-only hypothesis at the 5.1$\sigma$ level \cite{discovery}.
\vspace{-0.6cm}
\section{Description of the double-vertex event \label{sec::event}}
\begin{figure}
\centering
\includegraphics[width=.95\textwidth]{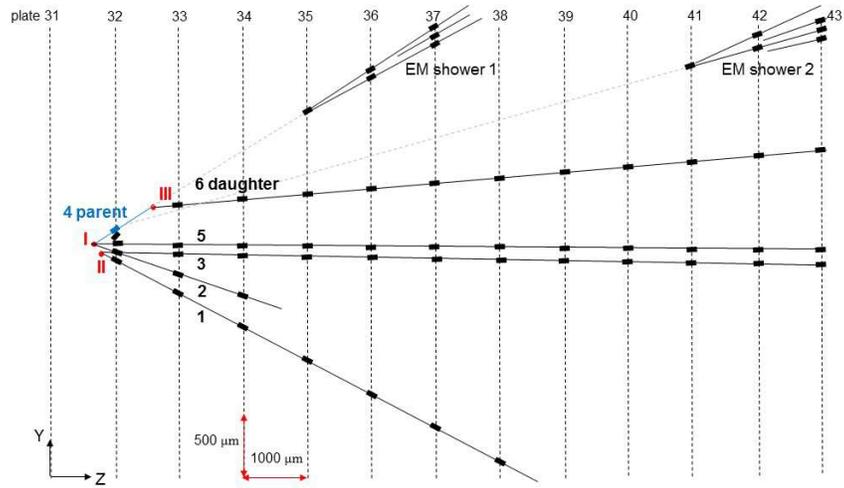}
\vspace{-0.6cm}
\caption{\small{Event projection in the $yz$ plane. Base-tracks are represented with segments, while volume tracks 
with lines; the blue one represents the daughter particle. Vertices are shown with red dots; EM showers 
1 and 2 are also indicated \label{img:event_side_view}} }
\end{figure}

\vspace{-0.2cm}
Event 1114301850 occurred in the first SM with no associated muon track 
(``0 $\mu$-event'' \cite{tau_papero}). The total hadronic energy is $\sim 20$ GeV. 
A scheme of the event is shown in Fig. \ref{img:event_side_view}. A primary stopping point was 
located in the lead layer between emulsion films 31 and 32, and a secondary one at plate 33. The upstream 
vertex is composed of 5 tracks; the downstream vertex has one prong kink topology. The five 
tracks are not compatible with a single vertex (IPs are larger than the standard threshold $\geq 10 \mu$m). 
By minimizing the IP the most probable event topology is composed of two vertices: a primary one ($V_I$) 
in \figurename~\ref{img:event_side_view} is $581.8~\mu$m upstream with respect to the top emulsion layer of 
plate 32; while the secondary one ($V_{II}$) is $102.6~\mu$m downstream with respect to $V_I$. $V_I$ 
is formed of tracks 2, 4 and 5, while $V_{II}$ is defined by tracks 1 and 3. IPs of all tracks with respect to 
both vertices are listed in Tab. \ref{tab:impact_parameters}. The distance between the primary and secondary 
vertices is $103.2~\mu$m.

Track 4 makes a ($95 \pm 17$) mrad kink ($537.6\pm 174$) $\mu$m upstream of plate 33; the minimum distance between
this track ``parent'' and track 6 ``daughter'' is ($0.9 \pm 2.9$) $\mu$m; the parent flight length 
is ($1174 \pm 74$) $\mu$m.  The kink point is labelled $V_{III}$. 
Track 2 is composed off 3 base-tracks (segments measured in three consecutive plates) and stops at plate 34; due 
to the short length its momentum, equal to ($0.31 \pm 0.08$) GeV$/c$, was evaluated considering absorption 
processes in lead \cite{pion_absorption_dependency}. Track 3 shows a re-interaction at plate 53 (not shown), 
while tracks 1 and 5 traverse the whole brick.
\begin{table}
\centering
\begin{tabular}{c|c|cc}
& \textbf{Single Vertex IP} ($\mu \mathrm{m}$) & \multicolumn{2}{c}{\textbf{Two vertices IP}} ($\mu \mathrm{m}$)\\
Track ID & & w.r.t. $V_{I}$ &  w.r.t. $V_{II}$\\
\hline
1 & 8.3  & 36.2 & \textbf{0.1}  \\
2 & 8.8  & \textbf{1.0}  & 6.5  \\
3 & 4.8  & 25.9 & \textbf{0.1}  \\
4 & 13.0 & \textbf{1.5}  & 20.4 \\
5 & 5.1  & \textbf{2.2}  & 9.6
\end{tabular}
\caption{\small{Impact Parameters evaluated assuming a topology of a single vertex,
or two vertices.
} \label{tab:impact_parameters} }
\end{table}
The momentum of tracks 1, 3, 5, and 6 was estimated by the MCS method, yielding $2.1^{+1.6}_{-3.1}$
GeV$/c$, $4.3^{+3.1}_{-7.1}$ GeV$/c$, $0.54^{+0.45}_{-0.68}$ GeV$/c$ and $2.7^{+2.1}_{-3.7}$ GeV$/c$ 
respectively (the quoted interval is the 68\% C.L.). 

Two electromagnetic (EM) showers were identified at plates 35 and 41. Their estimated energies are
$E_1 = $($7.2 \pm 1.7$) GeV and $E_2 = $($5.3 \pm 2.2)$ GeV, respectively. These photons could arise from 
one or two $\pi^0$. The two $\pi^0$ hypothesis is highly disfavoured because a second pair of EM showers was 
not found; by minimizing their IPs both photons are assumed emerging from vertex~$V_{III}$ ($8 \pm 8$) $\mu$m 
and ($40 \pm 11$)~$\mu$m.
An extensive search for nuclear fragments was performed and no fragments were detected in any of the three 
vertices. Both topology and kinematics were confirmed by measurements using an independent 
scanning system. 

In the OPERA proposal the observation probability of such a topology was taken as negligible, therefore no 
reconstruction procedure had been designed. Applying the standard selection criteria, 
none of the primary vertex tracks can be classified as a $\tau$ lepton. In particular the minimum daughter 
transverse momentum should be at least 300~MeV/c, while the best fit value for the parent is 242 MeV/c.
Processes which could produce two short decays (within $\sim$~1~mm) are: 
\begin{enumerate}
\vspace{-0.3cm}
\item $\nu_{\tau}$ CC interaction with charm production
\vspace{-0.3cm}
\item $\nu_{\mu}$ NC interaction with $c \overline{c}$ pair production. 
\vspace{-0.3cm}
\item other possibilities require a re-interaction of a final state particle in the lead; 
or a short decay of a semi-stable particle ($\pi$, $K$) within few millimetres from the 
primary vertex.
\vspace{-0.3cm}
\end{enumerate} 
Hadron re-interactions could mimic any of these possibilities:
\begin{itemize}
\vspace{-0.3cm}
\item $\nu_{\tau}$ CC interaction with hadron re-interaction
\vspace{-0.3cm}
\item $\nu_{\mu}$ NC interaction with two hadron re-interactions
\vspace{-0.3cm}
\item $\nu_{\mu}$ CC interaction with single charm production, with one hadron re-interaction and a misidentified muon
\vspace{-0.3cm}
\item $\nu_{\mu}$ CC interaction with two hadron re-interactions and a misidentified muon. 
\vspace{-0.3cm}
\end{itemize}
\vspace{-0.6cm}
\section{Analysis of the event}
\vspace{-0.2cm}
An analysis was performed using classifying algorithms to distinguish among decay
and hadronic interaction topologies, and therefore classify the observed event (section \ref{sec::event}).
The analysis is being finalized and here only classification results are presented without an estimate of 
the significance of the observation.

About 250 millions neutrino interaction events were generated using the GENIE \cite{genie_MC}, and
HERWIG~\cite{herwig_MC} (charm pair channel) generators. The simulation includes: $\nu_{\mu}$ CC DIS; $\nu_{\mu}$ 
NC DIS; and $\nu_{\tau}$ CC DIS interactions. All cases were simulated with and 
without charm production. The generation was done using fixed neutrino energies, then re-weighted considering 
the CNGS neutrino flux and the relative cross section of each process. Propagation through the OPERA brick was 
done with the GEANT4 framework \cite{geant4_1}, taking into account the experimental angular acceptance 
of reconstructed tracks in the OPERA emulsions and hadronic re-interactions inside the brick. Brick finding, 
vertex location, momentum reconstruction and muon identification efficiencies were taken into account using data 
driven parametrizations.

Only events with the same topology as the observed one were considered {\emph{i.e.}} no muon, electron or 
positron reconstructed at the primary vertex, one kink (one prong) secondary vertex with a charged daughter not 
reconstructed as a muon, and a two prong secondary vertex. Therefore, an event is defined 
as \emph{signal} when a $\tau$ plus a charm particle are selected by the topology cuts; while as \emph{background} 
in case of a double vertex topology without an identified $\tau$ particle. A set of twelve 
variables (topologic and kinematical) are used for the classification, namely: the daughter momentum 
and transverse momentum; the kink angle between parent and daughter; the charged and neutral parents flight length; 
the invariant mass; the total EM energy; the transverse angle $\varphi$ among tracks of the one and two prongs vertices; 
angle between charged and neutral parents; the missing transverse momentum at primary; and the hadronic momentum of tracks 
of the primary vertex, discarding tracks which belongs to a secondary vertex; the scalar sum of momentum of tracks of the 
neutral decay vertex.

\begin{figure}
\centering
\includegraphics[width=.7\textwidth]{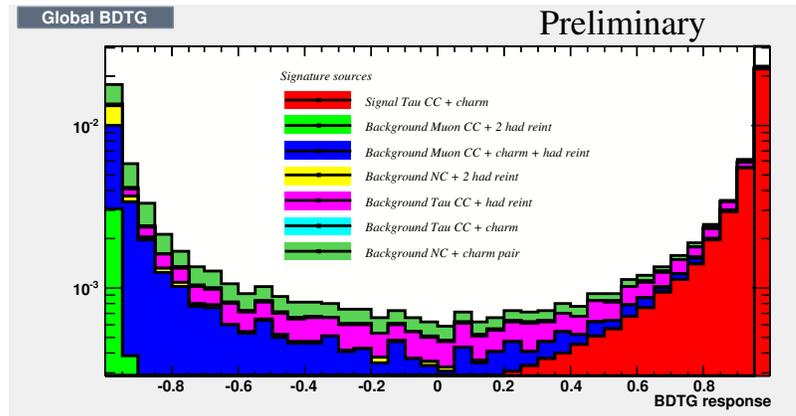}
\caption{ \small{BDTG classificator output, considering signal and several background-like channels. The vertical 
black line represents the preliminary BDTG classification of the measured event.}\label{fig:bdtg} }
\end{figure}

In the analysis three types of classifiers were used. A Multi Layer Perceptron flavour of an
Artificial Neural Network; 
two different kinds of a Boosted Decision Tree classifier, an Adaptive Boost and the Gradient Boost 
(BDTG). These are trees of binary choices taken on one single variable at a time until a stop 
criterion is fulfilled. With respect to the Adaptive Boost, the latter is more stable with respect to outliers which 
can occur by chance in any distribution, especially heavy-tailed distribution. Finally a Fisher discriminant was 
used \cite{NN}. \figurename~\ref{fig:bdtg} shows the preliminary output of the BDTG classifier, where the $\tau$ plus charm 
topology is peaked at 1; the contribution of all the other considered channels is also plotted. The black line is
the output evaluated using event 1114301850.
\vspace{-0.6cm}
\section{Conclusions}
\vspace{-0.2cm}
An event with two secondary vertices was measured by OPERA. A preliminary classification based in a multi-variate analysis, 
shows that this event turns to be likely a $\nu_{\tau}$ CC interaction with an associated charm production. 
Future perspectives of the analysis include the evaluation of the number of expected events for each considered
channel, and therefore obtain the significance of the observation. The analysis is in progress and a more 
comprehensive description will be given in a future publication which is in preparation. 
\vspace{-0.6cm}
\begin{small}

\end{small}

\end{document}